\documentclass{svmult}
\usepackage{epsfig,graphicx} 
\usepackage[bottom]{footmisc}
\usepackage{natbib,url}      
\bibpunct{(}{)}{;}{a}{}{,}   
\def\cite#1{\citealp{#1}}    
\def\authorindex#1{}
\def\figspath{.}  

\begin{document}\newcount\preprintheader\preprintheader=1


\def\thisvolume{these proceedings}

\def\aj{{AJ}}			
\def\araa{{ARA\&A}}		
\def\apj{{ApJ}}			
\def\apjl{{ApJ}}		
\def\apjs{{ApJS}}		
\def\ao{{Appl.\ Optics}} 
\def\apss{{Ap\&SS}}		
\def\aap{{A\&A}}		
\def\aapr{{A\&A~Rev.}}		
\def\aaps{{A\&AS}}		
\def\an{{Astron.\ Nachrichten}}
\def\aspcs{{ASP Conf.\ Ser.}}
\def\assp{{Astrophys.\ \& Space Sci.\ Procs., Springer, Heidelberg}}
\def\azh{{AZh}}			
\def\baas{{BAAS}}		
\def\jrasc{{JRASC}}	
\def\memras{{MmRAS}}		
\def\mnras{{MNRAS}}
\def\nat{{Nat}}		
\def\pra{{Phys.\ Rev.\ A}} 
\def\prb{{Phys.\ Rev.\ B}}		
\def\prc{{Phys.\ Rev.\ C}}		
\def\prd{{Phys.\ Rev.\ D}}		
\def\prl{{Phys.\ Rev.\ Lett.}} 
\def\pasp{{PASP}}
\def\pasj{{PASJ}}		
\def\qjras{{QJRAS}}
\def\science{{Sci}}		
\def\skytel{{S\&T}}		
\def\solphys{{Solar\ Phys.}} 
\def\sovast{{Soviet\ Ast.}}  
\def\ssr{{Space\ Sci.\ Rev.}}
\def\svassp{{Astrophys.\ Space Sci.\ Procs., Springer, Heidelberg}}
\def\zap{{ZAp}}			
\let\astap=\aap
\let\apjlett=\apjl
\let\apjsupp=\apjs
\def\grl{{Geophys.\ Res.\ Lett.}}  
\def\jgr{{J. Geophys.\ Res.}} 

\def\ion#1#2{{\rm #1}\,{\uppercase{#2}}}  
\def\deg{\hbox{$^\circ$}}
\def\sun{\hbox{$\odot$}}
\def\earth{\hbox{$\oplus$}}
\def\la{\mathrel{\hbox{\rlap{\hbox{\lower4pt\hbox{$\sim$}}}\hbox{$<$}}}}
\def\ga{\mathrel{\hbox{\rlap{\hbox{\lower4pt\hbox{$\sim$}}}\hbox{$>$}}}}
\def\sq{\hbox{\rlap{$\sqcap$}$\sqcup$}}
\def\arcmin{\hbox{$^\prime$}}
\def\arcsec{\hbox{$^{\prime\prime}$}}
\def\fd{\hbox{$.\!\!^{\rm d}$}}
\def\fh{\hbox{$.\!\!^{\rm h}$}}
\def\fm{\hbox{$.\!\!^{\rm m}$}}
\def\fs{\hbox{$.\!\!^{\rm s}$}}
\def\fdg{\hbox{$.\!\!^\circ$}}
\def\farcm{\hbox{$.\mkern-4mu^\prime$}}
\def\farcs{\hbox{$.\!\!^{\prime\prime}$}}
\def\fp{\hbox{$.\!\!^{\scriptscriptstyle\rm p}$}}
\def\micron{\hbox{$\mu$m}}
\def\onehalf{\hbox{$\,^1\!/_2$}}	
\def\onethird{\hbox{$\,^1\!/_3$}}
\def\twothirds{\hbox{$\,^2\!/_3$}}
\def\onequarter{\hbox{$\,^1\!/_4$}}
\def\threequarters{\hbox{$\,^3\!/_4$}}
\def\ubv{\hbox{$U\!BV$}}		
\def\ubvr{\hbox{$U\!BV\!R$}}		
\def\ubvri{\hbox{$U\!BV\!RI$}}		
\def\ubvrij{\hbox{$U\!BV\!RI\!J$}}		
\def\ubvrijh{\hbox{$U\!BV\!RI\!J\!H$}}		
\def\ubvrijhk{\hbox{$U\!BV\!RI\!J\!H\!K$}}		
\def\ub{\hbox{$U\!-\!B$}}		
\def\bv{\hbox{$B\!-\!V$}}		
\def\vr{\hbox{$V\!-\!R$}}		
\def\ur{\hbox{$U\!-\!R$}}


\def\labelitemi{{\bf --}}  

\def\rmit#1{{\it #1}}              
\def\rmit#1{{\rm #1}}              
\def\etal{\rmit{et al.}}           
\def\etc{\rmit{etc.}}           
\def\ie{\rmit{i.e.,}}              
\def\eg{\rmit{e.g.,}}              
\def\cf{cf.}                       
\def\viz{\rmit{viz.}}
\def\vs{\rmit{vs.}}

\def\rot{\hbox{\rm rot}}
\def\div{\hbox{\rm div}}
\def\lesssim{\mathrel{\hbox{\rlap{\hbox{\lower4pt\hbox{$\sim$}}}\hbox{$<$}}}}
\def\gtrsim{\mathrel{\hbox{\rlap{\hbox{\lower4pt\hbox{$\sim$}}}\hbox{$>$}}}}
\def\mathstacksym#1#2#3#4#5{\def#1{\mathrel{\hbox to 0pt{\lower 
    #5\hbox{#3}\hss} \raise #4\hbox{#2}}}}
\mathstacksym\lesssim{$<$}{$\sim$}{1.5pt}{3.5pt} 
\mathstacksym\gtrsim{$>$}{$\sim$}{1.5pt}{3.5pt} 
\mathstacksym\lrarrow{$\leftarrow$}{$\rightarrow$}{2pt}{1pt} 
\mathstacksym\lessgreat{$>$}{$<$}{3pt}{3pt} 

\def\dif{\: {\rm d}}                       
\def\ep{\:{\rm e}^}                        
\def\dash{\hbox{$\,-\,$}}                  
\def\is{\!=\!}                             

\def\starname#1#2{${#1}$\,{\rm {#2}}}  
\def\Teff{\hbox{$T_{\rm eff}$}}   

\def\kms{\hbox{km$\;$s$^{-1}$}}
\def\ms{\hbox{m$\;$s$^{-1}$}}
\def\Mxcm{\hbox{Mx\,cm$^{-2}$}}    

\def\Bapp{\hbox{$B_{\rm app}$}}    

\def\komega{($k, \omega$)}                 
\def\kf{($k_h,f$)}                         
\def\VminI{\hbox{$V\!\!-\!\!I$}}           
\def\IminI{\hbox{$I\!\!-\!\!I$}}           
\def\VminV{\hbox{$V\!\!-\!\!V$}}           
\def\Xt{\hbox{$X\!\!-\!t$}}                

\def\level #1 #2#3#4{$#1 \: ^{#2} \mbox{#3} ^{#4}$}   

\def\specchar#1{\uppercase{#1}}    
\def\AlI{\mbox{Al\,\specchar{i}}}  
\def\BI{\mbox{B\,\specchar{i}}} 
\def\BII{\mbox{B\,\specchar{ii}}}  
\def\BaI{\mbox{Ba\,\specchar{i}}}  
\def\BaII{\mbox{Ba\,\specchar{ii}}} 
\def\CI{\mbox{C\,\specchar{i}}} 
\def\CII{\mbox{C\,\specchar{ii}}} 
\def\CIII{\mbox{C\,\specchar{iii}}} 
\def\CIV{\mbox{C\,\specchar{iv}}} 
\def\CaI{\mbox{Ca\,\specchar{i}}} 
\def\CaII{\mbox{Ca\,\specchar{ii}}} 
\def\CaIII{\mbox{Ca\,\specchar{iii}}} 
\def\CoI{\mbox{Co\,\specchar{i}}} 
\def\CrI{\mbox{Cr\,\specchar{i}}} 
\def\CriI{\mbox{Cr\,\specchar{ii}}} 
\def\CsI{\mbox{Cs\,\specchar{i}}} 
\def\CsII{\mbox{Cs\,\specchar{ii}}} 
\def\CuI{\mbox{Cu\,\specchar{i}}} 
\def\FeI{\mbox{Fe\,\specchar{i}}} 
\def\FeII{\mbox{Fe\,\specchar{ii}}} 
\def\FeIX{\mbox{Fe\,\specchar{ix}}}
\def\FeX{\mbox{Fe\,\specchar{x}}}
\def\FeXVI{\mbox{Fe\,\specchar{xvi}}}
\def\FrI{\mbox{Fr\,\specchar{i}}}
\def\HI{\mbox{H\,\specchar{i}}} 
\def\HII{\mbox{H\,\specchar{ii}}} 
\def\Hmin{\hbox{\rmH$^{^{_{\scriptstyle -}}}$}}      
\def\Hemin{\hbox{{\rm He}$^{^{_{\scriptstyle -}}}$}} 
\def\HeI{\mbox{He\,\specchar{i}}} 
\def\HeII{\mbox{He\,\specchar{ii}}} 
\def\HeIII{\mbox{He\,\specchar{iii}}} 
\def\KI{\mbox{K\,\specchar{i}}} 
\def\KII{\mbox{K\,\specchar{ii}}} 
\def\KIII{\mbox{K\,\specchar{iii}}} 
\def\LiI{\mbox{Li\,\specchar{i}}} 
\def\LiII{\mbox{Li\,\specchar{ii}}} 
\def\LiIII{\mbox{Li\,\specchar{iii}}} 
\def\MgI{\mbox{Mg\,\specchar{i}}} 
\def\MgII{\mbox{Mg\,\specchar{ii}}} 
\def\MgIII{\mbox{Mg\,\specchar{iii}}} 
\def\MnI{\mbox{Mn\,\specchar{i}}} 
\def\NI{\mbox{N\,\specchar{i}}}
\def\NIV{\mbox{N\,\specchar{iv}}}
\def\NaI{\mbox{Na\,\specchar{i}}}
\def\NaII{\mbox{Na\,\specchar{ii}}}
\def\NaIII{\mbox{Na\,\specchar{iii}}}
\def\NeVIII{\mbox{Ne\,\specchar{viii}}} 
\def\NiI{\mbox{Ni\,\specchar{i}}} 
\def\NiII{\mbox{Ni\,\specchar{ii}}}
\def\NiIII{\mbox{Ni\,\specchar{iii}}} 
\def\OI{\mbox{O\,\specchar{i}}} 
\def\OVI{\mbox{O\,\specchar{vi}}}
\def\RbI{\mbox{Rb\,\specchar{i}}} 
\def\SII{\mbox{S\,\specchar{ii}}} 
\def\SiI{\mbox{Si\,\specchar{i}}} 
\def\SiII{\mbox{Si\,\specchar{ii}}} 
\def\SrI{\mbox{Sr\,\specchar{i}}}
\def\SrII{\mbox{Sr\,\specchar{ii}}}
\def\TiI{\mbox{Ti\,\specchar{i}}} 
\def\TiII{\mbox{Ti\,\specchar{ii}}} 
\def\TiIII{\mbox{Ti\,\specchar{iii}}} 
\def\TiIV{\mbox{Ti\,\specchar{iv}}} 
\def\VI{\mbox{V\,\specchar{i}}} 
\def\HtwoO{\mbox{H$_2$O}}        
\def\Otwo{\mbox{O$_2$}}          

\def\Halpha{\mbox{H\hspace{0.1ex}$\alpha$}} 
\def\Ha{\mbox{H\hspace{0.2ex}$\alpha$}}
\def\Hbeta{\mbox{H\hspace{0.2ex}$\beta$}}
\def\Hgamma{\mbox{H\hspace{0.2ex}$\gamma$}}
\def\Hdelta{\mbox{H\hspace{0.2ex}$\delta$}}
\def\Hepsilon{\mbox{H\hspace{0.2ex}$\epsilon$}}
\def\Hzeta{\mbox{H\hspace{0.2ex}$\zeta$}}
\def\Lyalpha{\mbox{Ly$\hspace{0.2ex}\alpha$}}
\def\Lybeta{\mbox{Ly$\hspace{0.2ex}\beta$}}
\def\Lygamma{\mbox{Ly$\hspace{0.2ex}\gamma$}}
\def\Lycont{\mbox{Ly\hspace{0.2ex}{\small cont}}}
\def\Baalpha{\mbox{Ba$\hspace{0.2ex}\alpha$}}
\def\Babeta{\mbox{Ba$\hspace{0.2ex}\beta$}}
\def\Bacont{\mbox{Ba\hspace{0.2ex}{\small cont}}}
\def\Paalpha{\mbox{Pa$\hspace{0.2ex}\alpha$}}
\def\Bralpha{\mbox{Br$\hspace{0.2ex}\alpha$}}

\def\NaD{\mbox{Na\,\specchar{i}\,D}}    
\def\NaDone{\mbox{Na\,\specchar{i}\,\,D$_1$}}
\def\NaDtwo{\mbox{Na\,\specchar{i}\,\,D$_2$}}
\def\NaID{\mbox{Na\,\specchar{i}\,\,D}}
\def\NaIDone{\mbox{Na\,\specchar{i}\,\,D$_1$}}
\def\NaIDtwo{\mbox{Na\,\specchar{i}\,\,D$_2$}}
\def\Done{\mbox{D$_1$}}
\def\Dtwo{\mbox{D$_2$}}

\def\Mgbone{\mbox{Mg\,\specchar{i}\,b$_1$}}
\def\Mgbtwo{\mbox{Mg\,\specchar{i}\,b$_2$}}
\def\Mgbthree{\mbox{Mg\,\specchar{i}\,b$_3$}}
\def\MgIb{\mbox{Mg\,\specchar{i}\,b}}
\def\MgIbone{\mbox{Mg\,\specchar{i}\,b$_1$}}
\def\MgIbtwo{\mbox{Mg\,\specchar{i}\,b$_2$}}
\def\MgIbthree{\mbox{Mg\,\specchar{i}\,b$_3$}}

\def\CaIIK{\mbox{Ca\,\specchar{ii}\,K}}       
\def\CaIIH{\mbox{Ca\,\specchar{ii}\,H}}
\def\CaIIHK{\mbox{Ca\,\specchar{ii}\,H\,\&\,K}}
\def\HK{\mbox{H\,\&\,K}}
\def\Kthree{\mbox{K$_3$}}      
\def\Hthree{\mbox{H$_3$}}
\def\Ktwo{\mbox{K$_2$}}
\def\Htwo{\mbox{H$_2$}}
\def\Kone{\mbox{K$_1$}}     
\def\Hone{\mbox{H$_1$}}     
\def\KtwoV{\mbox{K$_{2V}$}}
\def\KtwoR{\mbox{K$_{2R}$}}
\def\KoneV{\mbox{K$_{1V}$}}
\def\KoneR{\mbox{K$_{1R}$}}
\def\HtwoV{\mbox{H$_{2V}$}}
\def\HtwoR{\mbox{H$_{2R}$}}
\def\HoneV{\mbox{H$_{1V}$}}
\def\HoneR{\mbox{H$_{1R}$}}

\def\hk{\mbox{h\,\&\,k}}
\def\kthree{\mbox{k$_3$}}    
\def\hthree{\mbox{h$_3$}}
\def\ktwo{\mbox{k$_2$}}
\def\htwo{\mbox{h$_2$}}
\def\kone{\mbox{k$_1$}}     
\def\hone{\mbox{h$_1$}}     
\def\ktwoV{\mbox{k$_{2V}$}}
\def\ktwoR{\mbox{k$_{2R}$}}
\def\koneV{\mbox{k$_{1V}$}}
\def\koneR{\mbox{k$_{1R}$}}
\def\htwoV{\mbox{h$_{2V}$}}
\def\htwoR{\mbox{h$_{2R}$}}
\def\honeV{\mbox{h$_{1V}$}}
\def\honeR{\mbox{h$_{1R}$}}

\ifnum\preprintheader=1     
\makeatletter  
\def\@maketitle{\newpage
\markboth{}{}%
  {\mbox{} \vspace*{-8ex} \par 
   \em \footnotesize To appear in ``Magnetic Coupling between the Interior 
       and the Atmosphere of the Sun'', eds. S.~S.~Hasan and R.~J.~Rutten, 
       Astrophysics and Space Science Proceedings, Springer-Verlag, 
       Heidelberg, Berlin, 2009.} \vspace*{-5ex} \par
 \def\lastand{\ifnum\value{@inst}=2\relax
                 \unskip{} \andname\
              \else
                 \unskip \lastandname\
              \fi}%
 \def\and{\stepcounter{@auth}\relax
          \ifnum\value{@auth}=\value{@inst}%
             \lastand
          \else
             \unskip,
          \fi}%
  \raggedright
 {\Large \bfseries\boldmath
  \pretolerance=10000
  \let\\=\newline
  \raggedright
  \hyphenpenalty \@M
  \interlinepenalty \@M
  \if@numart
     \chap@hangfrom{}
  \else
     \chap@hangfrom{\thechapter\thechapterend\hskip\betweenumberspace}
  \fi
  \ignorespaces
  \@title \par}\vskip .8cm
\if!\@subtitle!\else {\large \bfseries\boldmath
  \vskip -.65cm
  \pretolerance=10000
  \@subtitle \par}\vskip .8cm\fi
 \setbox0=\vbox{\setcounter{@auth}{1}\def\and{\stepcounter{@auth}}%
 \def\thanks##1{}\@author}%
 \global\value{@inst}=\value{@auth}%
 \global\value{auco}=\value{@auth}%
 \setcounter{@auth}{1}%
{\lineskip .5em
\noindent\ignorespaces
\@author\vskip.35cm}
 {\small\institutename\par}
 \ifdim\pagetotal>157\p@
     \vskip 11\p@
 \else
     \@tempdima=168\p@\advance\@tempdima by-\pagetotal
     \vskip\@tempdima
 \fi
}
\makeatother     
\fi

\title*{Probability Density Functions to Represent Magnetic Fields
at the Solar Surface}
\titlerunning{Probability Density Functions}       
\author{M. Sampoorna}
\authorindex{Sampoorna, M.}
\institute{Indian Institute of Astrophysics, Koramangala,
Bangalore 560 034, India}
\maketitle

\setcounter{footnote}{0}

\begin{abstract}
Numerical simulations of magneto-convection and analysis of solar
magnetogram data provide empirical probability density functions
(PDFs) for the line-of-sight component of the magnetic field. In this
paper, we theoretically explore effects of several types of PDFs on
polarized Zeeman line formation. We also propose composite PDFs to
account for randomness in both field strength and orientation. Such
PDFs can possibly mimic random fields at the solar surface.
\end{abstract}

\section{Introduction}
\label{sampoorna-sec:intro}
Magneto-convection on the Sun has a size spectrum that spans several
orders of magnitude and develops eddies with sizes much smaller than
the spatial resolution of current spectro-polarimeters (about 0.2
arcsec or 150 km at the photospheric level).
Thus the Stokes profiles that we observe are always averages over
space, time, and along the line-of-sight (LOS). This suggests that it would 
be appropriate to characterize the magnetic field responsible for
spectral line polarization by a probability density function (PDF).
The case of Gaussian PDFs was studied by
\citet{sampoorna-1972SvA....16..450D,sampoorna-1979Ap&SS..66...47D,sampoorna-2005A&A...442...11F,sampoorna-2006ASPC..358..126F,sampoorna-2006A&A...453.1095F,sampoorna-2007MmSAI..78..142F,sampoorna-2008NewA...13..233S}.
In this paper we study PDFs determined from observations by
\citet{sampoorna-2002ESASP.505..101S,sampoorna-2003ASPC..286..169S,sampoorna-2003AN....324..397S} and from 
magneto-convection simulations by \citet[][ ~see also \cite{sampoorna-2005A&A...429..335V}]{sampoorna-2006ApJ...642.1246S}.\  
\citet{sampoorna-2008A&A...485..275S} used these PDFs to compute mean 
solutions in the macro and micro-turbulent limits. Macro and micro-turbulence are 
also referred to as optically thick and optically thin limits. In this paper we 
discuss in detail the mean solutions computed for a more general regime of meso-turbulence. 

It is well known that photospheric photon mean free paths (50--100~km in 
the optical) correspond approximately to the sizes of meso-turbulent magnetic 
eddies. Calculation of mean Stokes parameters
in this regime was considered by\ 
\citet{sampoorna-1994ssm..work...29L,sampoorna-2006A&A...453.1095F,sampoorna-2006ASPC..358..126F,
sampoorna-2003AN....324..392C,sampoorna-2005AN....326..296C,sampoorna-2007A&A...468..323C}. We
use a Kubo-Anderson Process (KAP) with correlation length $1/\nu$
(where $\nu$ is number of jumps per unit optical depth) and a PDF to
characterize the random magnetic field\
\citep{sampoorna-2006A&A...453.1095F}. Using KAP in a Milne-Eddington
model atmosphere,\ \citet{sampoorna-2006A&A...453.1095F} deduce
explicit expressions for the mean and rms fluctuations of the emergent
Stokes parameters.  In this paper we use those expressions to compute
mean solutions for magnetic eddies of arbitrary size.

\section{Scalar PDFs}
\label{sampoorna-scalar-pdf}
Recently\
\citet{sampoorna-2002ESASP.505..101S,sampoorna-2003ASPC..286..169S,sampoorna-2003AN....324..397S}
have found from an analysis of high resolution La Palma and MDI solar
magnetograms that the PDF for the LOS component is nearly independent
of the spatial scale and can be well represented by a Voigt
function. This PDF has a Gaussian core (centered around zero field)
with a magnetic width $\Delta_B$, and Lorentzian wings with a magnetic
damping parameter $a_B$. The quantity $\Delta_B$ is a measure of the
rms fluctuations of the LOS component.

A Voigt PDF with $\Delta_B=6\,$G and $a_B=1.5$ best fits the empirical
PDF for LOS field strength derived from La Palma magnetograms. A
symmetric Voigt function has zero net flux. In real magnetograms
magnetic flux is locally unbalanced. Non-zero net flux can be
generated by a Voigt PDF that has a symmetric core, but different
$a_B$ for positive and negative polarities.

Magneto-convection simulation near the 
solar surface by\ \citet{sampoorna-2006ApJ...642.1246S} show that the magnetic field is 
intermittent with a stretched exponential distribution. Here  
we restrict ourselves to asymmetric scalar PDFs as they represent the 
photospheric conditions better. 

\subsection{Asymmetric Voigt PDF}
\label{sampoorna-asymm_voigt}
In terms of parameters $a_B$ and $\Delta_B$, symmetric Voigt PDF has 
the form 
\begin{equation}
P_{\rm V} (y,a_B) = {\frac{a_B}{\pi^{3/2}}}
\,\int_{-\infty}^{+\infty}\,{\frac{e^{-u^2}}{(y-u)
^2 + a^2_B}}\,{\rm d}u.
\label{sampoorna-voigt-pdf-nondim}
\end{equation}
In the above equation we have introduced non-dimensional parameters 
\begin{equation}
y = {B}/{\Delta_B},\quad u = {B_1}/{\Delta_B},\quad
\gamma_B = {\Delta_B}/{B_{\rm D}},   
\label{sampoorna-definition}
\end{equation}
where ${1}/{B_{\rm D}} = g{e}/({4\pi m c}{\Delta \nu_{\rm D}})$, 
in standard notation. 
Here $B$ is the random magnetic field component along a given direction,
and $\Delta \nu_{\rm D}$ is the frequency 
Doppler width. Thus $\gamma_B$ represents rms fluctuations $\Delta_B$ 
converted to Zeeman shift in Doppler width units.   

\begin{figure}
\centering
\includegraphics[height=5cm,width=11.5cm]{\figspath/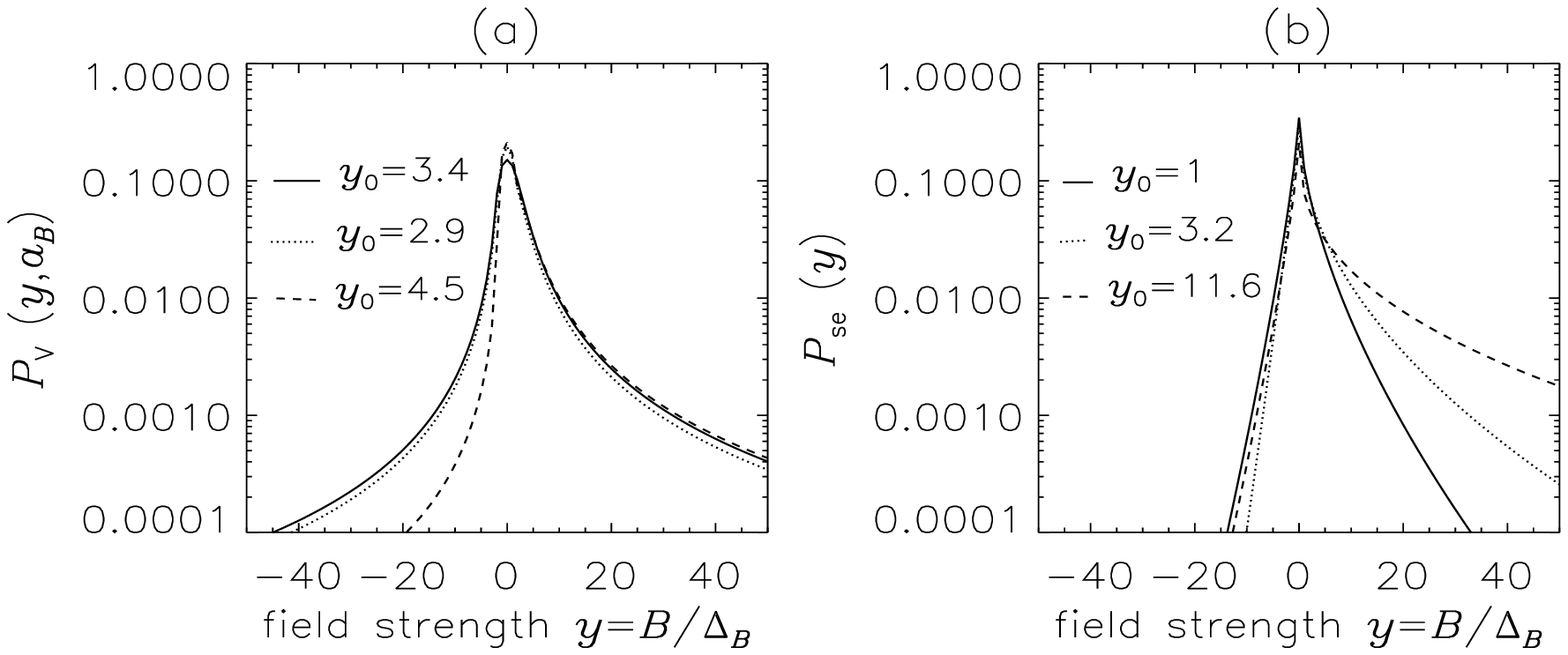}
\caption[]{\label{sampoorna-asy_voigt}
(a) Asymmetric Voigt and (b) Stretched exponential PDFs. The 
  $y$-scale is related to the $B$-scale through 
 $\Delta_B$\ 
  \citep[$=6$ G according to][]{sampoorna-2002ESASP.505..101S}. 
}  
\end{figure}

Asymmetric Voigt PDFs can be constructed by choosing different
values of $a_B$ for different parts of
the PDF while keeping the Gaussian core symmetrical.
Figure~\ref{sampoorna-asy_voigt}a shows examples of asymmetric PDFs\ 
\citep[see\,][\,for\ details]{sampoorna-2008A&A...485..275S}, which more or less resemble
the PDF for the La Palma magnetogram shown in Fig.~2 of\
\citet{sampoorna-2002ESASP.505..101S}.  The mean magnetic field $y_0$ is
the average of $y$ over $P_{\rm V}(y,a_B)$.

Mean emergent residual Stokes parameters $\langle r_{I,Q,U,V}\rangle$
are computed using asymmetric scalar PDFs for magnetic eddies of
arbitrary size.  The random field has a fixed orientation with respect
to the LOS defined by polar angles $(\gamma,\,\chi) = (60^{\circ},\,
30^{\circ})$. Emergent residual Stokes parameters are defined by
$r_{I} = [I_c - I]/C_1$ and $r_{X} = -{X}/C_1$, with the symbol $X$
denoting $Q$, $U$ or $V$. The constant $C_1$ is the slope of continuum
source function $S(\tau_c) = C_0 + C_1 \tau_c$, with $\tau_c$ the
continuum optical depth.  The continuum intensity at the surface is
$I_c = C_0+C_1$.  This model has damping parameter $a=0$ and line
strength parameter (ratio of line to continuum absorption coefficient)
$\beta = k_0/k_{\rm c}=10$. We also assume that the spectral line has
a wavelength around 5000\,\AA, a Land\'e factor of 2 and a Doppler
width of 1.5\,km\,s$^{-1}$. For this typical line, $B_{\rm
D}=1.07\times 10^{3}$ G. Hence $\gamma_B=0.0056$ for an rms magnetic
field fluctuation $\Delta_B=6\,$G. As a result the $\langle r_I
\rangle$ profiles in all figures in this paper remain insensitive to
the PDF parameters.

\begin{figure}
\centering
\includegraphics[height=8.0cm,width=11.5cm]{\figspath/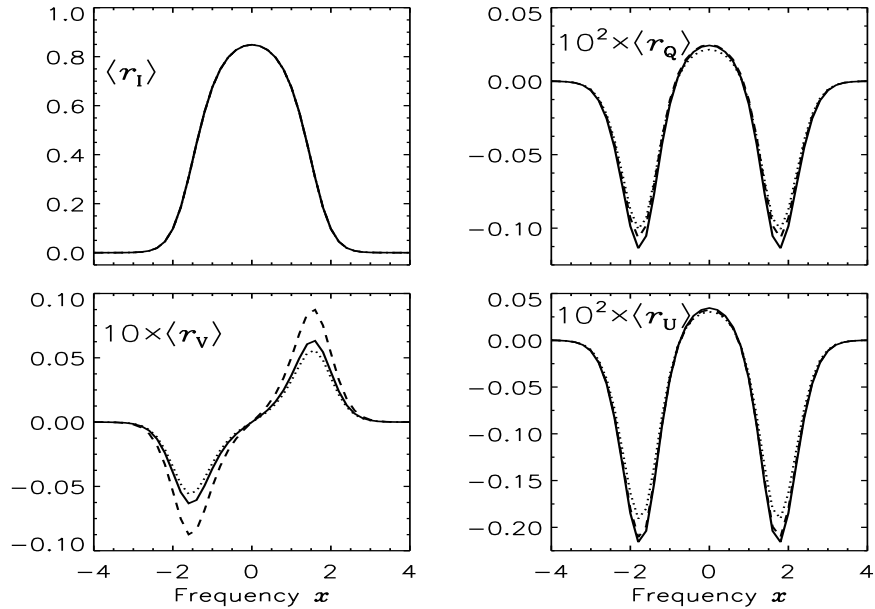}
\caption[]{\label{sampoorna-asy_voigt_emerg}
$\langle r_{I,Q,U,V}\rangle$ computed with asymmetric Voigt 
PDFs in Fig.~\ref{sampoorna-asy_voigt}a for $\nu=5$. Reference to 
line types is the same as in Fig.~\ref{sampoorna-asy_voigt}a. }
\end{figure}

Figure~\ref{sampoorna-asy_voigt_emerg} shows $\langle r_{I,Q,U,V}\rangle$ 
computed using the 3 PDFs in Fig.~\ref{sampoorna-asy_voigt}a for $\nu=5$. 
Note that $\langle r_V\rangle$ is zero if one uses a symmetric PDF with   
zero mean field. The $\langle
r_{Q,U}\rangle$ profiles show very small sensitivity to 
PDF asymmetry, while the $\langle r_I\rangle $ profiles indeed are
insensitive.  For 
all three PDFs, $\langle r_{V}\rangle$ peaks around $x \approx 1.5$. 
The peak amplitudes increase with the mean field. 

\subsection{Asymmetric stretched exponential PDF}
\label{sampoorna-strexppdf}
A symmetric stretched exponential may be written in functional form as
\begin{equation}
\label{sampoorna-str-exp-pdf}
 P_{\rm se} (y)\,{\rm d}y = C\,{\rm e}^{-|y|^k}\,{\rm d}y.
\end{equation}
The quantity $k$ takes values 
between 0 and 1 and is referred to as the stretching parameter. $C$ is
the normalization constant.

We construct asymmetric stretched exponential PDFs with non-zero mean field 
by choosing different $k$ values
  for positive and negative polarities. Figure~\ref{sampoorna-asy_voigt}b
  shows three examples\ 
\citep[see\,][\,for\ details]{sampoorna-2008A&A...485..275S}. 
\begin{figure}
\centering
\includegraphics[height=8cm,width=11.5cm]{\figspath/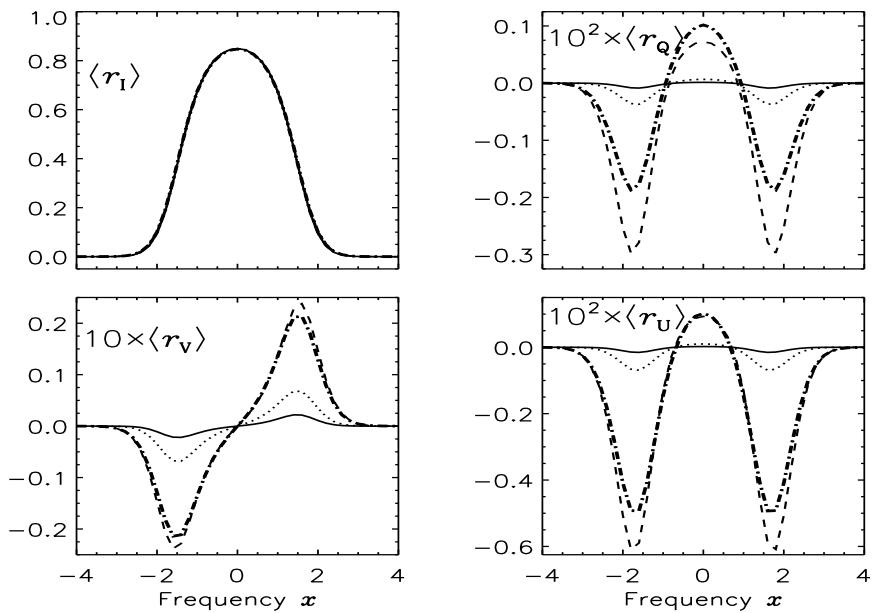}
\caption[]{\label{sampoorna-asy_str-exp-riquv}
$\langle r_{I,Q,U,V}\rangle$ for
asymmetric stretched exponential PDFs. Solid,
  dotted and dashed lines refer to $\nu=5$ (the same line types as in 
  Fig.~\ref{sampoorna-asy_voigt}b). Heavy dot-dashed line refers to 
macro-turbulent limit for $y_0=11.6$.}
\end{figure}

Mean residual Stokes profiles computed with these three PDFs are shown
in Fig.~\ref{sampoorna-asy_str-exp-riquv}.  For $\nu=5$, we observe a
clear increase in the peak amplitudes of $\langle r_{Q,U,V}\rangle$
when the mean field $y_0$ increases. However, the peak positions are
essentially insensitive to $y_0$. The $\langle r_I\rangle$ profiles
remain insensitive to PDF asymmetry.  Differences between
the solutions for $\nu=5$ and the macro-turbulent limit appear for $\langle
r_{Q,U}\rangle$ when $y_0=11.6$, due to the extended tail of the
PDF for positive polarities. For $\langle r_V\rangle$, the differences
remain small even for $y_0=11.6$. The relative insensitivity of
$\langle r_V\rangle$ to the scale of magnetic field fluctuations has been
observed for Voigt type PDFs\
\citep[see][]{sampoorna-2008A&A...485..275S} and also for symmetric
Gaussian PDFs with non-zero mean field\
\citep{sampoorna-2006A&A...453.1095F}. It is due to the fact that
the PDF asymmetry is sharply peaked around $y_0$.  In the limit of
a Dirac distribution, there would be no difference between solutions
for different $\nu$ values since the magnetic field would be
deterministic. It thus appears that a mean value of Stokes $V$ can be
calculated with reasonable confidence using the micro-turbulent limit, a
remark made already in
\citet{sampoorna-2006A&A...453.1095F}. 

\section{Angular PDFs}
\label{sampoorna-joshfpl}
A large fraction of the solar atmosphere is filled with mixed polarity
fields, and the inter-granular lanes contain fields that are
preferentially directed upward or downward. To represent this situation, 
we consider magnetic fields that have a fixed value of the
strength $B$ but random orientations. For such a random field,
following angular distribution has been suggested by\
\citet{sampoorna-1987SoPh..114....1S}\,:
\begin{equation}
P_{\rm pl}(\mu_B) = {\frac{(p+1)}{4\pi}}\,|\mu_B|^p, \quad 
-1 \le \mu_B \le +1. 
\label{sampoorna-pl-pdf}
\end{equation}
Here, $\mu_B = \cos\theta_B$, with $\theta_B$ the field orientation
with respect to the atmospheric normal.  The power law index $p$ can
take any value; the $p=0$ case corresponds to an isotropic
distribution. As $p$ increases the distribution becomes more and more
peaked in the vertical direction.
\begin{figure}
\centering
\includegraphics[height=5cm,width=11.5cm]{\figspath/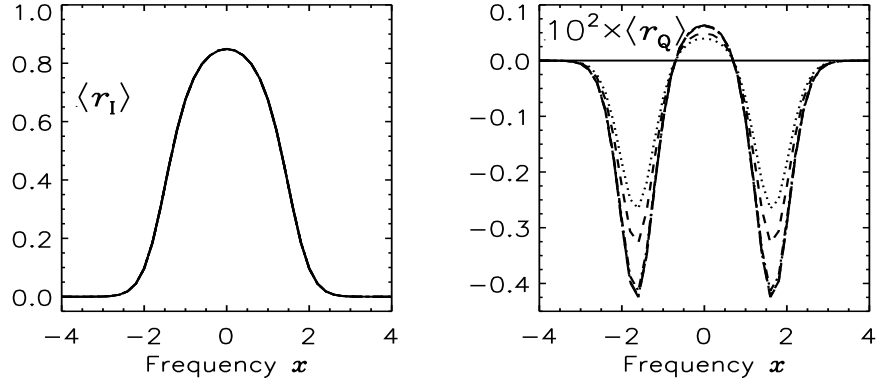}
\caption[]{\label{sampoorna-fig-plhf1} 
$\langle r_{I,Q} \rangle$ at $\mu=0.1$ (limb observation)
computed for $\nu=5$ and angular power law PDF 
  (Eq.~(\ref{sampoorna-pl-pdf})).    
Line types\,: 
$p=0$ (solid), 5 (dotted), 10 (dashed), 100 (dot-dashed), 500 
(dash-triple-dotted), and 1000 (long dashed).  
In this case $\langle r_{U,V} \rangle =0$.}  
\end{figure}

Mean profiles $\langle r_{I,Q} \rangle$ are calculated for $\nu=5$,
magnetic field strength $B/B_{\rm D}= 0.1$, line strength $\beta =
10$, damping parameter $a=0$, and the power law index $p$ a free
parameter.  Figure~\ref{sampoorna-fig-plhf1} shows $\langle r_{I,Q}
\rangle$ at the limb ($\mu=0.1$). A comparison of this figure with
Fig.~12 of\ \citet{sampoorna-2008A&A...485..275S} computed for
the micro-turbulent limit shows that there is very little difference
between the two. This is because the absolute value of
the magnetic field along the LOS is bounded by the condition $B/B_{\rm
D}=0.1$. Since the field is weak, $\langle r_Q \rangle \ll \langle r_I
\rangle$. Further, $\langle r_{Q}\rangle$ is zero for $p=0$. As $p$
gets larger, $\langle r_{Q}\rangle$ first increases with $p$ and then
saturates for $p\simeq 100$ (see\ Fig.~\ref{sampoorna-fig-plhf1}).

\section{Vector magnetic field distributions}
\label{sampoorna-combpdf}
To describe a random vector magnetic field we need a PDF that
simultaneously accounts for strength and angle fluctuations.  {\it
From physical considerations one may argue that angular variation
should be strongly field-strength dependent}\
\citep[see\,][]{sampoorna-2008arXiv0812.4465D}. For the strongest
fields the distribution should be peaked around the vertical direction, as
strong fields tend to have intermittent fluxtube morphology.  The
weakest fields on the other hand would be passively moved and bent by
turbulent fluid motions and so get so tangled up that their distribution
would be nearly isotropic. The transition from isotropic to peaked
distributions would probably be gradual (possibly around 50\,G).

\begin{figure}
\centering
\includegraphics[height=8cm,width=11.5cm]{\figspath/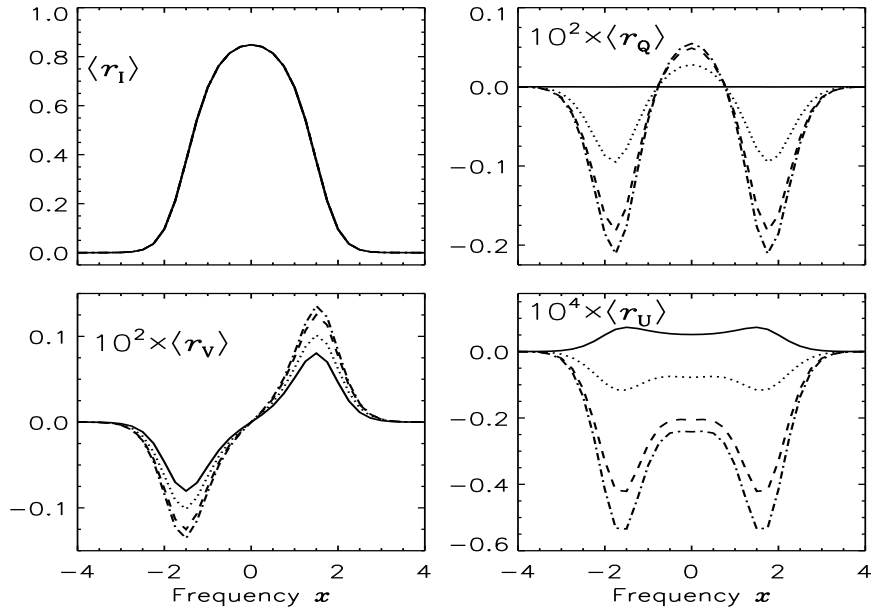}
\caption[]{\label{sampoorna-Vpl-rimu1}
$\langle r_{I,Q,U,V}\rangle$ for $\nu=5$ 
and $\mu=0.1$. Composite PDF containing asymmetric Voigt
function with $y_0=4.5$ (dashed line of
Fig.~\ref{sampoorna-asy_voigt}a) is used. Line types 
are\,: $y_t=\infty$ (solid), $y_t=50$ (dotted), $y_t=10$
(dashed), and $y_t=5$ (dash-dotted).  }
\end{figure}

Based on this scenario we propose PDFs here which are combinations of
angle and field strength distributions. For the angular part we use
the power law distribution introduced in Sect.~\ref{sampoorna-joshfpl},
while for the field strength part we consider either a Voigt function
(see\ Sect.~\ref{sampoorna-asymm_voigt}) or a stretched exponential
(see\ Sect.~\ref{sampoorna-strexppdf}).  The functional form of such a
composite PDF is
\begin{eqnarray}
P(y,\mu_B,\phi_B)\,{\rm d}\mu_B\,{\rm d}\phi_B\,{\rm d}y = 
{\frac{(p+1)}{2\pi}}\,
\cases{
P_{\rm V}(y, a_B)\,
\mu^p_B\, {\rm d}\mu_B\,{\rm d}\phi_B\,{\rm d}y,  \cr
C\,e^{-|y|^k} \,
\mu^p_B\, {\rm d}\mu_B\,{\rm d}\phi_B\,{\rm d}y.} 
\end{eqnarray}
Here $y$ varies in the range $[-y_{\rm max}, +y_{\rm max}]$, $\mu_B$
in the range $[0,1]$ and $\phi_B$ in the range $[0,2\pi]$.  $P_{\rm
V}(y, a_B)$ is given in
Eq.~(\ref{sampoorna-voigt-pdf-nondim}). Asymmetric composite PDFs can
be constructed by choosing different $a_B$ (or $k$) values for
positive and negative polarities of the Voigt (or stretched
exponential).  The angle and strength distributions are coupled by
letting the power law index $p$ depend on $y$. We have chosen $p
=|y|/y_t$ with $y_t = B_t / \Delta_B$, where $B_t$ marks the
transition between isotropic and peaked.  Note that $y_t=\infty$
corresponds to fully isotropic distribution for all field strengths.

We calculate mean residual Stokes parameters for the composite PDF
containing an asymmetric Voigt with mean field $y_0=4.5$ (dashed curve in
Fig.~\ref{sampoorna-asy_voigt}a).  The model parameters are
$(a,\beta,\gamma_B) = (0,10,0.0056)$ and
$\mu=0.1$. Figure~\ref{sampoorna-Vpl-rimu1} shows the solutions for
$\nu=5$. The different curve types correspond to different $y_t$
values.  $\langle r_I\rangle$ is insensitive to the PDF asymmetry 
and to the variation of $y_t$ due to the very weak value of
$\gamma_B$.  When $y_t\to\infty$, the PDF becomes fully isotropic, and
hence $\langle r_Q
\rangle\to 0$. As $y_t$ decreases, the PDF becomes more and more
anisotropic and hence $\langle r_Q \rangle$ as well as $\langle r_V
\rangle$ increase in magnitude. In this case $\langle r_U \rangle$ is 
generated through the magneto-optical effects. Therefore
$\langle r_U\rangle$ is very small, with a behavior similar to
$\langle r_Q \rangle$.

\section{Conclusions}
\label{sampoorna-conclu}
We have presented mean Stokes profiles formed in media having
spatially unresolved magnetic structures with sizes that are
comparable to photon mean free paths, using
PDFs that describe fluctuations of the ambient field. A Gaussian
PDF with isotropic or anisotropic fluctuations was considered in\
\citet{sampoorna-2005A&A...442...11F}; here, we experimented with other
types of PDFs with restriction to asymmetric PDFs which can
generate non-zero net flux as diagnosed by the
shape of $\langle r_V \rangle$ profile.  We consider very weak
fluctuations of the magnetic field ($\gamma_B=0.0056$).  Thus $\langle
r_{I} \rangle$ profiles remain insensitive to the shape of PDF. In
contrast, other mean Stokes profiles are quite sensitive to the choice
of PDF.

For a complete description of the random vector magnetic field we need
PDFs which describe both the angular and strength fluctuations.  We
constructed such empirical PDFs by combining a power law (for angular
distribution) with a Voigt function or a stretched exponential (for
field strength). At the solar surface, weak fields are observed to
be nearly isotropic and strong fields more vertical.  Composite PDFs
can simulate such a situation.

\begin{acknowledgement}
I am grateful to Dr. K.~N. Nagendra for very useful suggestions and 
comments. 
\end{acknowledgement}
\begin{small}

\end{small}
\end{document}